\documentclass[vecphys]{svmult}     
\usepackage{graphicx}        
\usepackage{epsfig}          
\usepackage{subfigure}
\begin{document}

\title*{Multidimensional Solitons -- Theory}
\author{
L.~D.~Carr\inst{1} \and J.~Brand\inst{2}}

\institute{Department of Physics, Colorado School of Mines, Golden,
CO
  80401, U.S.A.;\\Electron and Optical Physics Division, National
  Institute of Standards and Technology, Gaithersburg, MD 20899-8410
  U.S.A. \texttt{lcarr@mines.edu} \and Centre of Theoretical Chemistry
  and Physics, Institute of Fundamental Sciences, Massey University
  Auckland, New Zealand.  \texttt{J.Brand@massey.ac.nz}}
\maketitle

\section{Introduction}

The one-dimensional solitons described in Chapters II and III of
this book can be extended into two and three dimensions.  Such
extensions are generally  unstable~\cite{sulem1999}. However, in the
tightly confined geometries associated with trapped Bose-Einstein
condensates (BECs) both bright and dark solitons extended into two
and three dimensions can be stabilized for times longer than the
lifetime of experiments~\cite{muryshev99,carr2000e,feder2000}. BECs
offer one the opportunity to tune a matter-wave gradually from one
to two to three dimensions~\cite{dalfovo1999,leggett2001}.  In the
crossover regimes, new nonlinear objects can even appear, such as
the \emph{svortex}, a solitary wave which is a soliton-vortex
hybrid~\cite{brand2001,brand2002}.  The general question of
crossover dimensions is an intriguing one in physics, as real
experimental systems never have a perfect integer number of
dimensions.

We will consider both the stable and unstable regimes of higher
dimensional solitons, treating such objects theoretically but with
an eye towards BEC experiments.  BECs are typically contained in
harmonic traps, and have a profile ranging from Gaussian to inverse
parabolic, depending on the interaction strength~\cite{dalfovo1999}.
They span tens to hundreds of microns.  Their lifetime is on the
order of one to a hundred seconds.  Both thermal and quantum
fluctuations can play a significant role in their
dynamics~\cite{dalfovo1999}.  We must take into account all of these
factors when discussing solitons.  Moreover, the finite non-uniform
nature of trapped BECs leads to significantly different nonlinear
dynamics than that found in the GPE for uniform media.  To cite a
simple example, even in one dimension with periodic boundary
conditions the finite domain of the condensate leads to spontaneous
symmetry breaking and quantum phase
transitions~\cite{carr2000b,kanamoto2005}.

There are also solitons which do not have any 1D analog.  For
instance, a vortex-anti-vortex pair in two dimensions is a solitary
wave as it represents a localized excitation which moves
coherently~\cite{jones1982}.  An example in three dimensions is a
vortex ring~\cite{donnelly1991,saffman1992,anderson2001}. More
complicated topological solitons, such as skyrmions, are possible in
multi-component condensates~\cite{mueller2004}.  We will discuss
such solitons as well.

In keeping with the theme of this book, we will deal mathematically
only with the mean field theory of BECs, described by the
Gross-Pitaevskii Equation (GPE), and linear perturbation of the mean
field, described by the Boguliubov-de-Gennes equations (BDGE). The
GPE and BDGE can be derived rigorously from first principles from a
second quantized quantum field theory for binary interactions
between atoms in a dilute weakly interacting Bose gas well below the
critical temperature for Bose-Einstein condensation, as discussed in
Chapter I and the references therein.  We note that there are
significant subtleties in interpretation of BDGE solutions; see the
appendix of Ref.~\cite{garay2001} for a discussion of these issues.
The GPE plus BDGE picture has an excellent interpretation in terms
of quantum fluid dynamics, as discussed by Fetter and
Svidzinsky~\cite{fetter2001}.

Lastly, we note that the majority of higher-dimensional results,
particularly for non-uniform trapped BECs, are achieved numerically.
Many excellent references in computational science describe rigorous
numerical methods for the GPE and BDGE (e.g.~\cite{perezgarcia2003}
and references therein).  Due to the paucity of exact analytical
results, we focus primarily on a coherent summary of numerical
studies.

We introduce a small set of notation before proceeding.  The
effective nonlinearity is given by $g_{\mathrm{eff}}=g N$, which can
be obtained by a simple rescaling of the wavefunction amplitude.
Then the wavefunction $\Psi(\vec{r},t)$ is normalized to unity. An
axisymmetric harmonic trap can be characterized by its asymmetry
parameter $\lambda\equiv\omega_z/\omega_r$, where $\omega_r$ is the
radial trapping frequency and $\omega_z$ the axial trapping
frequency.  The harmonic oscillator lengths are given by
$\ell_z\equiv \sqrt{\hbar/m\omega_z}$ and $\ell_r\equiv
\sqrt{\hbar/m\omega_r}$ .

\section{Dark Solitons and Solitary Waves in Higher Dimensions}

\subsection{Dark Band and Planar Solitons}

\begin{figure}[p]
     \centering
     \subfigure[Snake instability in a spherical trap]{
          \label{fig:1a}
          \includegraphics[width=0.95\textwidth,height=9.0cm]{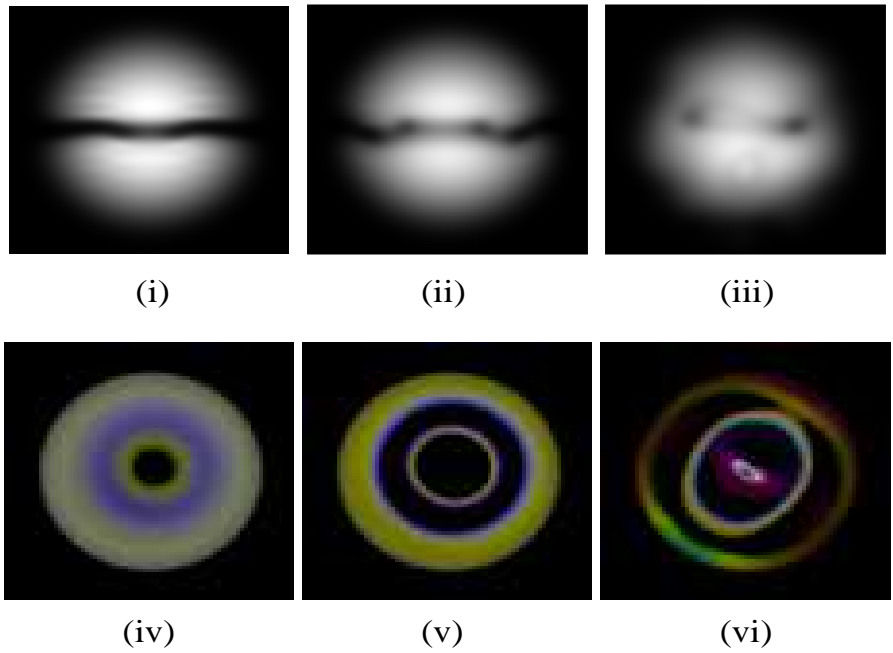}}\\
     \vspace{.1in}
     \subfigure[Snake instability in a non-axisymmetric trap]{
           \label{fig:1b}
           \includegraphics[width=0.95\textwidth,height=8.0cm]{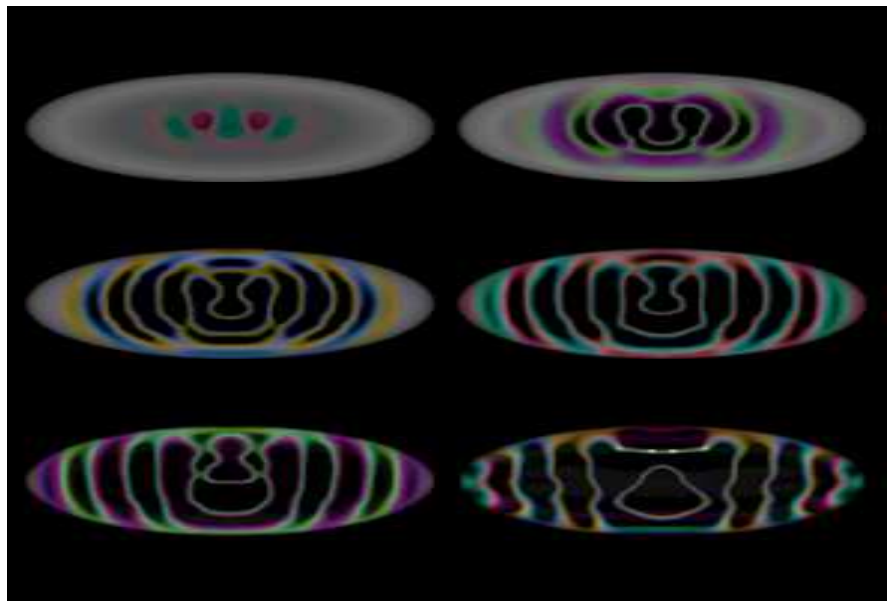}}
     \caption{Dynamical instability of a single planar soliton in
     a trapped Bose-Einstein condensates.
     See text for full description.}
     \label{fig:1}
\end{figure}

In a three dimensional system a standing dark soliton takes the form
of a planar node; in two dimensions the node is a line, sometimes
called a band.  When such a soliton moves with respect to the
background condensate, the notch fills in, so that the density in
the region of the soliton is reduced but does not form a node; a
detailed description is provided in Chapter III.  We term these
\emph{planar solitons} and \emph{band solitons}, respectively. In
uniform media it is well known that both planar and band solitons
decay via the snake instability.  A sinusoidal mode transverse to
the plane/band grows exponentially. The arcs of this ``snake'' break
off into vortex-anti-vortex pairs. In the context of Bose-Einstein
condensates, this has actually been suggested as a way to produce
both vortices and anti-vortices in the same condensate, which is not
otherwise possible with the usual stirring techniques.

The decay time for band solitons has been calculated with the
BDGE, i.e., by considering linear perturbations to band/planar dark
soliton stationary state of the GPE~\cite{law1993,mcdonald1993}.
\emph{Nonlinear} instability times can be significantly shorter, and
are determined from the numerical integration of the GPE.  A band
soliton then decays into an infinite chain of vortex anti-vortex
pairs. Depending on the initial condition, these pairs can join to
form solitary waves or annihilate in vortex-anti-vortex
collisions.

The nonlinear dynamics following decay of a planar soliton can be
significantly more complex, as linear excitations leading to the
snake instability can occur in two dimensions.  The vortices
produced are vortex lines which can rotate and/or combine to form
vortex rings~\cite{komineas2007}.  Keeping in mind that a velocity
field can equally well be characterized by a vorticity field under
certain simple assumptions~\cite{saffman1992}, the decay of large
arrays of planar vortices can lead to turbulence, characterized by
densely tangled vortex lines~\cite{schwarz1985,schwarz1988}.

In trapped BECs the situation is quite different. In the
three-dimensional harmonic trap the condensate, for sufficiently
large nonlinearity, has a central parabolic profile and Gaussian
tails.  In this regime, called the \emph{Thomas-Fermi regime}, as
defined in Chapter I, there is an additional mechanism for
instability.  Since a planar soliton moves at a fraction of the
sound velocity which depends on its depth, and the sound velocity is
proportional to the square root of density~\cite{dalfovo1999}, the
non-uniform density profile causes the soliton to travel more slowly
at the edges of the condensate than the center. An initially uniform
planar soliton formed in the center of the trap deforms into a
U-shaped propagation front.  When this wavefront reaches the edge of
the trap it is deflected, and the trailing edges curl up to form
vortices~\cite{denschlag2000}.  Thus there are two competing
instability mechanisms for planar dark solitons in BECs. In initial
experiments on planar solitons, it was in fact the
non-uniformity-induced instability which dominated, as discussed in
Chapter Vb.  Here, we emphasize the snake instability.

Shown in Fig.~\ref{fig:1} (reproduced with permission of the
authors~\cite{feder2000}) are the precise dynamics of the snake
instability in a harmonic trap for an initially stationary planar
soliton with realistic experimental parameters. In Fig.~\ref{fig:1a}
the condensate contains $10^5$ atoms and is in a spherical trap with
$\omega_r=\omega_z=2\pi\times 50$~rad/s. The initial single planar
soliton state is obtained with imaginary time relaxation.  Shown in
the panels are snapshots after real time propagation of 47~ms,
50~ms, and 77~ms for Fig.~\ref{fig:1a}(i)-(iii) and (iv)-(vi). In
(i)-(iii), the brightness is proportional to the condensate density,
and the images correspond to densities integrated down the line of
sight. In (iv)-(vi), the brightness is {\it inversely} proportional
to the condensate density, and regions outside the Thomas-Fermi
sphere are rendered transparent in order to visualize nodes in the
condensate interior; the color corresponds to the phase: $\phi=0$
through $2\pi$ is represented by the sequence red-green-blue-red.
The view is perpendicular to the original nodal plane of the
soliton; prior to the snake instability the dark soliton would
appear as a featureless disk.

In Fig.~\ref{fig:1b} the breakup of an initial planar soliton is
shown as a function of time for $N=10^6$ atoms, $\omega_x=2\pi\times
14$~rad/s, $\omega_y/\omega_x=\sqrt{2}$, and $\omega_z/\omega_x=2$,
the precise geometry of Ref.~\cite{denschlag2000}.  From the top
left to the bottom right in raster order are shown times $t=15$~ms
through $20$~ms in 1~ms increments after the initial state is
formed. The view is along $\hat{y}$, and the Hamiltonian was
constrained to even parity along $\hat{x}$ and $\hat{z}$ for ease of
computation. The rendering is identical to that of
Figs.~\ref{fig:1a}(iv)-(vi). The filamentation is almost entirely
constrained to the original nodal $xz$-plane.

These figures describe only the mean field picture.  Recent studies
have shown that finite temperature can cause significant dissipative
effects even in one dimension~\cite{burger1999,jackson2006}, while
coupling to transverse modes can lead to dissipation within the
GPE/BDGE picture~\cite{fedichev1999}.   Moreover, even in one
dimension quantum fluctuations determined by the BDGE ``blur'' a
dark soliton, due to uncertainty in the position of the density
minimum~\cite{dziarmaga2003}.  A full theory of dark band and planar
soliton dynamics with even lowest order finite temperature and
quantum effects remains a significant challenge to the computational
and theoretical scientific communities, although general theoretical
prescriptions in this direction
exist~\cite{zaremba1999,morgan1999,davis2002a}.

\subsection{Ring Dark Solitons and Spherical Shell Solitons}
\label{sec:RDS}

Another way to create a higher dimensional soliton is to wrap a band
or planar soliton back around on itself.  In two dimensions this
takes the form of a nodal ring, termed a \emph{dark ring soliton}.
In three dimensions such an object is a nodal spherical shell,
termed a \emph{spherical shell soliton}.  These objects are always
unstable in harmonic traps, but can have lifetimes longer than that
of BEC experiments.  Multiple ring solitons can be nested within
each other.  It is mathematically intriguing that such solutions are
\emph{nonlinear Bessel functions}, by which we mean solutions to the
equation
\begin{equation} \eta_q''+\frac{1}{\chi}\eta_q'-\frac{q^2}{\chi^2}\eta_q -
\eta_q^3 +\eta_q=0 \, ,\label{eqn:gpe2d}\end{equation} where the
wavefunction has been rescaled as
\begin{equation} \psi(\vec{r},t)=\sqrt{\frac{\mu}{g}}\,\eta_q(\chi)\exp(i q \phi)\exp(-i \mu
t/\hbar)\exp(i\theta_0)\,. \label{eqn:assume}\end{equation} Here
$\mu$ is the chemical potential, $q$ is the winding number of a
central vortex, $\theta_0$ is an arbitrary phase, and the coordinate
system is cylindrical with coordinates $\chi,\phi$, with
$\chi\equiv(\sqrt{2m \mu}/\hbar)r$. Equation~(\ref{eqn:gpe2d}) is
clearly the generating equation of a Bessel function, modified by
the nonlinear term $\eta_q^3$.

\begin{figure}[t]
\center
\includegraphics[width=0.95\textwidth]{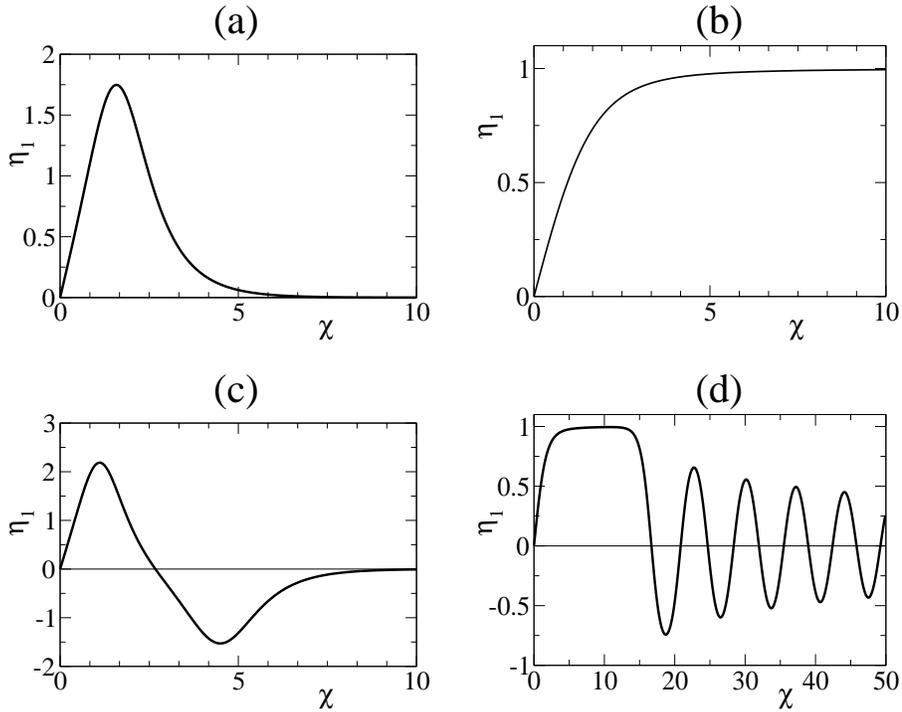}
\caption{{\it  A quantum vortex of winding number $q=1$ in free
space}: (a) attractive case; (b) repulsive case.  It is evident in
(a) that a bright vortex is also a bright ring soliton.  A radially
excited state: (c) the first excited state in the attractive case;
(d) in the repulsive case, a radially excited state requires an
infinite number of nodes and asymptotically resembles the Coulomb
function~\cite{carr2006d}. The radial dependence of the order
parameter of an infinitely extended condensate is depicted. Note
that all axes are dimensionless: $\eta_1$ is a rescaled radial
density while $\chi$ is a rescaled radial coordinate.}
\label{fig:2}       
\end{figure}

In infinitely extended repulsive condensates, the solutions to
Eq.~(\ref{eqn:assume}) include the uniform ground state, singly and
multiply-quantized vortices, and ring soliton solutions, as
illustrated in Fig.~\ref{fig:2}(b) and Fig.~\ref{fig:2}(d).  The
latter require a countably infinite number of nested dark ring
solitons, where each soliton is a radial node corresponding to a
node of the nonlinear Bessel function.  The asymptotic form of these
solutions has been studied in Ref.~\cite{carr2006d}.

In contrast, in harmonically trapped condensates, ring solitons can
be added to a solution one by one, so that there is a denumerably
infinite set of ring soliton solutions for fixed nonlinearity.  The
linear stability analysis of these solutions and subsequent
nonlinear dynamics of their breakup has been studied via the BDGE
and GPE in the context of both
BECs~\cite{theocharis2003,theocharis2005,carr2006d} and
optics~\cite{kivshar1994,dreischuh1996,neshev1997,frantzeskakis2000,nistazakis2001,dreischuh2002}.
The dominant decay modes of single ring solitons in harmonically
trapped BECs with and without a central vortex of winding number
unity are the quadrupole and octupole,
respectively~\cite{carr2006d}.  In general, instabilities in higher
dimensions can lead to new nonlinear structures; in
Refs.~\cite{theocharis2003,theocharis2005} it is shown that ring
solitons decay into vortex necklaces, as was later observed in
optics experiments~\cite{yangJ2005}.
Refs.~\cite{theocharis2003,theocharis2005} also provide an
analytical description of ring soliton dynamics.  It has been
suggested that, by use of an optical phase-shifting technique such
as that employed in creating planar
solitons~\cite{denschlag2000,burger1999,carr2001e}, one might be
able to generate ring solitons in experiments on BECs and observe
their subsequent dynamics.

We note that structures similar to those of the stationary spherical
shell solitons have been observed as transients in an experiment by
Ginsberg {\em et al.} \cite{ginsberg2005} and in simulated
collisions of vortex rings \cite{komineas:110401} as will be
discussed in Sec.~\ref{sec:solitary} and Chapter Vb.

\subsection{Solitary Waves in Restricted Geometries} \label{sec:solitary}
\newcommand{\nn}{n_1}
\newcommand{\vs}{v_s}
\renewcommand{\d}{{\mathrm d}}

When a condensate is confined to a strongly prolate harmonic trap
such that its transverse dimensions are not so small as to approach
the healing length but not so large as to be effectively
three-dimensional, new families of solitary waves arise.  It
suffices to consider the case $\omega_z=0$, $\omega_r\neq 0$, so
that the condensate forms an infinitely long cylinder.  Physically,
one can loosely interpret this as a multi-mode waveguide, where a
uniform condensate forms the ``vacuum'' and solitary waves can
propagate in the $z$ direction.  In this picture, one maintains a
finite linear particle density $n_1 = N/L$, counting the number of
particles per unit length along the symmetry axis $z$.

A mathematical representation of this geometry is realized by the
GPE in the following dimensionless form:
\begin{equation}  \label{eq:gp}
i\, \frac{\partial\Psi}{\partial t} = -\frac{1}{2}\, \Delta \Psi +\frac{1}{2}\,r^2
\,\Psi + 4\pi\gamma\,|\Psi|^2 \Psi,
\end{equation}
where $r=\sqrt{x^2+y^2}$ is the radial coordinate and $\Delta =
\partial^2/\partial x^2 +\partial^2/\partial y^2 + \partial^2/\partial
z^2$ is the Laplacian operator~\cite{komineas02}. The
dimensionless coupling constant $\gamma\equiv\nn a$ is the only
parameter entering the equation, where $a$ is the scattering length.
Length is measured in units of the transverse oscillator length
$\ell_r$ and the unit of time is $1/\omega_r$.  At $z \to \pm \infty$
the wave function approaches the ground state in the transverse
plane with $\partial\Psi/\partial z=0$ and a transverse
normalization of $2\pi \int{dr\,r |\Psi|^2} = 1$.

\begin{figure}[ht]
\center
\includegraphics[width=0.8\textwidth,viewport=58 124 535
308,clip]{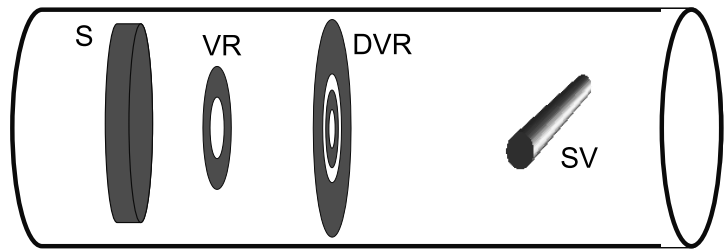} \caption{Schematic of solitary wave
configurations in a cylindrical
  BEC. The band soliton (S) has a nodal plane perpendicular to the
  trap axis and shows axial and inversion symmetry. The vortex ring
  (VR) is axisymmetric and has a vortex line (line singularity in the
  phase) in the configuration of a closed ring around the trap axis.
  The double vortex ring (DVR) has the same symmetries as the VR but
  features two concentric loops of vortex lines. The solitonic vortex
  (SV) has no symmetry. A vortex line is configured perpendicular to
  the trap axis and does not close in itself but terminates at the
  condensate boundaries.}\label{fig:3a}
\end{figure}

The parameter $\gamma$ characterizes the dimensionality of the
problem. In fact, $\gamma$ is closely related to the number of
healing lengths that fit into the transverse diameter of the
cylindrical BEC cloud. Consequently, $\gamma \ll 1$ is the
one-dimensional regime, where the transverse profile of the density
is Gaussian and the waveguide is single-mode.  In this regime, the
only solitary waves known are the familiar family of dark solitons
from the one-dimensional NLS. Three-dimensional aspects  only become
relevant when we consider effects that are sensitive to the breaking
of integrability in the system as it has been found, e.g., in the
interactions of phonons with solitons \cite{Muryshev2002a}. For
$\gamma \gg 1$, the condensate enters the Thomas-Fermi regime for
which the transverse density profile is approximated by an inverted
parabola. In this regime various families of solitary waves with
different structures co-exist. Figure \ref{fig:3a} shows a schematic
of possible configurations.

\begin{figure}[ht]
\center
\parbox{0.68\textwidth}{
\includegraphics[width=0.65\textwidth]{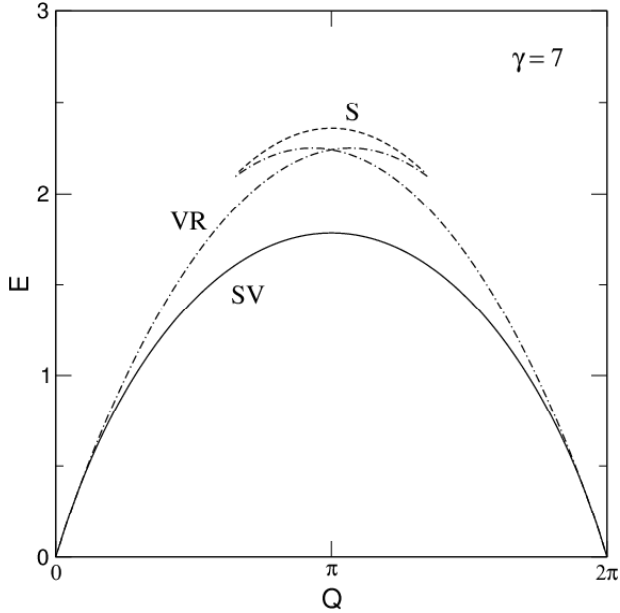}}
\parbox{0.3\textwidth}{
\caption{The energy $E$ versus impulse $Q$ dispersion relation of the
  known solitary waves in the cylindrical BEC at $\gamma=7$ from
  Ref.~\cite{komineas:043617}. Shown are the branches of the svortex
  (SV), vortex rings (VR), and the band soliton (S). The slope $dE/dQ$
  of the dispersion relation gives the velocity of the solitary wave.
  The density structure of vortex rings and the time-dynamics of their
  head-on collision is shown in Fig.~\ref{fig:4}.}\label{fig:3}}
\end{figure}

Families of cylindrically symmetric solitary waves have been
numerically characterized by Komineas and
Papanicolaou~\cite{komineas02,komineas:023615,komineas2007} and
consist of dark solitons, vortex rings and double rings (see
Fig.~\ref{fig:3a}). The only non-axisymmetric solitary wave
discovered so far is the solitonic vortex, or
svortex~\cite{brand2001}, which consists of a vortex line
perpendicular to the cylinder axis.  This nonlinear excitation has
solitonic properties in that it is a stable solitary wave which
propagates coherently, and can be generated by stirring in toroidal
traps or by spontaneous decay from an unstable band
soliton~\cite{brand2001,brand2002}.  The dispersion relation of the
svortex was calculated for a cylindrical geometry in
Ref.~\cite{komineas:043617}, as shown in Fig.~\ref{fig:3}
(reproduced with permission of the authors~\cite{komineas:043617}).

The picture that emerges from the numerical calculations is the
following.  For $\gamma < 1.5$ the situation is
quasi-one-dimensional and only one solitary wave with the essential
properties and structure of the 1D dark soliton exists. For
$\gamma>1.5$ there is a bifurcation and the non-axisymmetrix svortex
excitation coexists with axisymmetric solitary waves, i.e., band
solitons. For $\gamma>4$, there is another bifurcation and vortex
rings coexist with band solitons and svortices. For even larger
$\gamma$ more bifurcations can be expected leading to a ``zoo'' of
solitary waves. Numerical calculations up to $\gamma=20$ have been
performed in Refs.~\cite{komineas:023615,komineas:043617}. The
stability properties of these families of solitary waves have not
been studied in detail, although svortices and vortex rings are
believed to be dynamically stable.

\begin{figure}[t]
\center
\includegraphics[width=0.95\textwidth]{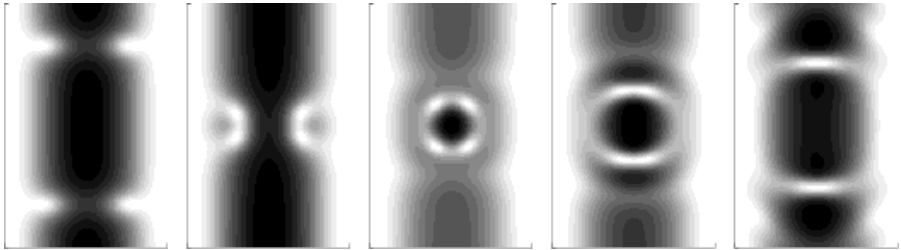}
\caption{Inelastic collision of vortex rings produces a transient
shell
  structure of near spherical symmetry, from
  Ref.~\cite{komineas:110401}. Plotted is the density $|\Psi|^2$ on the
  $y=0$ plane at different time frames showing the head-on collision of a
  pair of vortex rings at $\gamma=7$ with $v \equiv dE/dQ = 0.34 v_s$,
  where $v_s = 1.61$ is the speed of sound in the units of Eq.~(\ref{eq:gp}).\label{fig:4}}
\end{figure}

The stability of vortex rings in particular has been tested
numerically by simulating head-on collisions~\cite{komineas:110401},
as shown in Fig~\ref{fig:4}.  It was found that vortex rings collide
elastically at large and small velocities while dramatically violent
collisions occur at intermediate velocities.  While these results
could be explained in terms of the known dispersion diagrams like
Fig.~\ref{fig:3}, a peculiar observation from the simulations was
that inelastic collisions can generate shell structures of nearly
spherical symmetry reminiscent of the spherical shell solitons
discussed in Sec.~\ref{sec:RDS}. Similar structures were also
observed in the experiment of Ref.~\cite{ginsberg2005}, which will
be discussed in more detail in Chap. Vb.

\subsection{Vortex Rings and Rarefaction Pulses}

\begin{figure}[t]
\center
\includegraphics[width=0.95\textwidth]{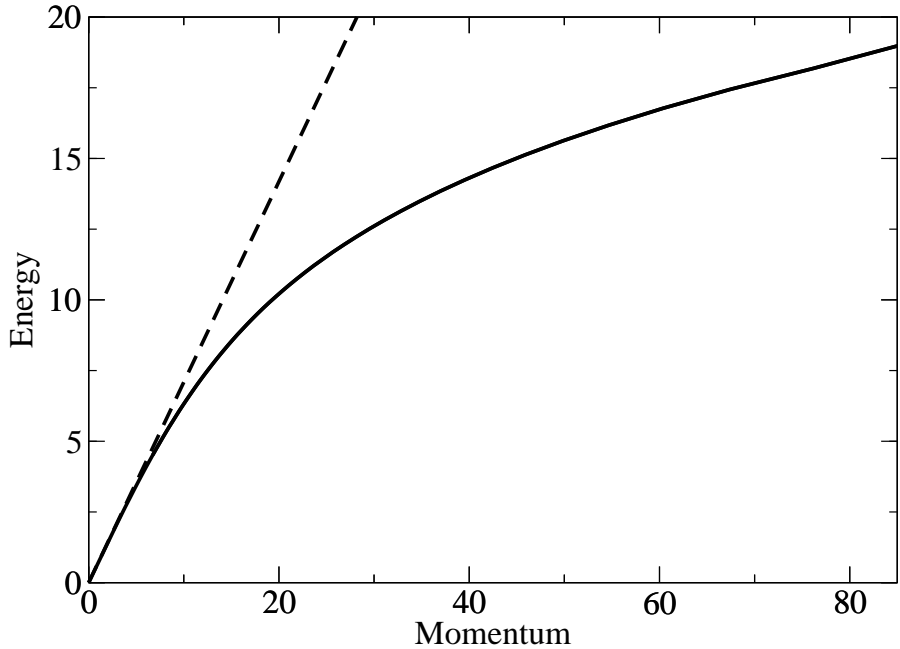}
\caption{Dispersion relation~\cite{jones1982,noteBerloff} for 2D
($xz$) solitary waves (solid line). The dashed line shows the
dispersion relation of sound for comparison.}
\label{fig:5}       
\end{figure}

We now turn to the discussion of solitary waves that are localized
on a significantly smaller scale than the condensate dimensions. In
this case the idealization to a homogeneous condensate is
appropriate, and the local density approximation can be applied to
obtain non-uniform results.  In fact, most of the work done in this
direction assumes an infinite and homogeneous background density.
The ground-breaking work in this area was done in the 1980's by
Jones and Roberts, who computed the dispersion relation of 2D and 3D
solitary waves~\cite{jones1982}, as reproduced in
Fig.~\ref{fig:5}~\cite{noteBerloff}. In three dimensions, the
solitary waves they found are a family of vortex rings with varying
diameter, which is related to the energy, the impulse, and the
velocity.  As the diameter is decreased to the order of the healing
length, the phase singularity disappears; nevertheless, a branch of
solitary waves can be found. As these waves have reduced particle
density in the region where they are localized, they are also called
{\em rarefaction pulses}. Analytical formulas in the form of
Pad\`{e} approximants for 3D solitary waves are given in
Ref.~\cite{berloff04a}.

The stability of the Jones-Roberts solitons was first discussed in
Ref.~\cite{jones1986}, but rigorous results were only obtained
recently \cite{berloff2004,tsuchiya06:2Dsolitons}. It is interesting
to note that simulations showed that head-on collisions of vortex
rings are always highly inelastic \cite{koplik96}, in contrast to the
situation in cylindrical traps discussed above. There has also been
some work on the interactions between 3D solitons, vortex lines, and
phonon radiation in the context of superfluid turbulence
\cite{leadbeater03,berloff2004}. Recent work in a hydrodynamic
framework hints that the properties
of vortex rings can be dramatically modified by Kelvin-wave
excitations to the extent that the vortex rings may change their direction
of propagation \cite{barenghi:046303}.

Another specific sort of multidimensional soliton is a solitary wave
moving along a vortex line that extends through a homogeneous
condensate. Such excitations were discussed recently in
Ref.~\cite{berloff:010403}.

\subsection{Multi-component Bose-Einstein condensates}

Multi-component condensates offer rich opportunities for the study
of solitonic and solitary waves in higher dimensions.  A great deal
of work has been done in this area.  We touch very briefly on this
subject.  The main idea behind topological solitons in
multi-component BECs is as follows.  Repulsive inter-species and
intra-species interactions in a multi-component BEC will tend to
make the total particle density uniform by filling up low-density
regions of vortex cores with particles from another component. Under
the assumption of a uniform total density, the vectorial order
parameter has a prescribed constant length and becomes a mapping of
three-dimensional real space to a sphere.  Topological solitons are
found as solutions with nontrivial topology resulting from this
mapping.  Examples of topological solitons are skyrmions in three
dimensions and baby-skyrmions in two dimensions. A very large number
of different vortex textures are possible in multi-component
BECs~\cite{mueller2004,ruostekoski2004,kasamatsu2005,pogosov2005},
Various suggestions have been made to observe skyrmions and other
such objects in BECs, but none have been achieved so far in
experiments~\cite{busch99,khawaja01,PhysRevA.64.043612,savage:010403,battye02prl,herbut:080403,PhysRevLett.86.3934,mizushima02}.
There is an ongoing discussion about the potential stability and
experimental observability of such solutions~\cite{zhai:043602}. In
the case of dipolar BECs where long-range interactions play a role
in addition to the contact interaction considered so far, spin
textures may form spontaneously and the stability conditions change
\cite{yi:020401}.

We would also like to mention that a generalization of the
Jones-Roberts solitons to two-component condensates with a variety
of different solitary wave families is described in
Ref.~\cite{berloff:120401}. Nonlinear phenomena in multi-component
BECs are discussed in more detail in Chapter X.

\section{Bright Solitons in Higher Dimensions}

While dark solitons are excitations of a condensate that take the
form of density notches, bright solitons are ground or metastable
states of a condensate, even in higher dimensions.  Thus an
attractive BEC is \emph{itself} a soliton.  Most experiments have
focused on the unstable regimes of bright
solitons~\cite{sackett1998,sackett1999,donley2001}; we will also
discuss the many theoretical proposals based on stable regimes,
which are only just beginning to be explored in
experiments~\cite{carr2002b,strecker2002,cornish2006}.

\subsection{Instability, Metastability, Stability}
\label{ssec:ims}

\begin{figure}[t]
\center
\includegraphics[width=0.95\textwidth,height=8.5cm]{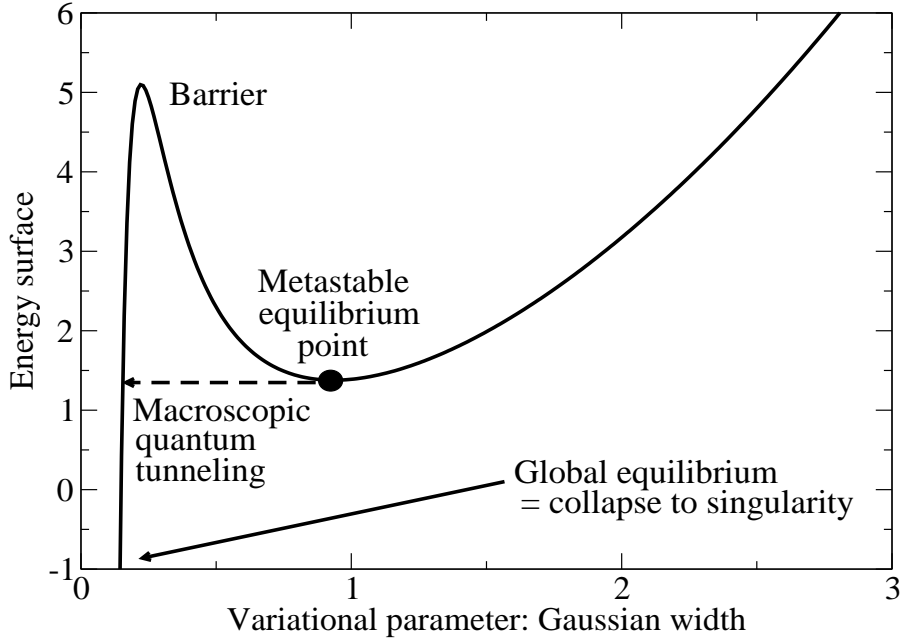}
\caption{{\it Metastability and macroscopic quantum tunneling:}
Shown is a representative energy surface for a variational Gaussian
ansatz in an isotropic 3D harmonic trap for attractive interactions
$g_{\mathrm{eff}}<0$, as a function of the single variational
parameter, the width of the Gaussian.  One observes that
non-singular solutions (non-zero width) are always metastable.  The
height of the collapse barrier decreases as $N\rightarrow N_c^-$;
$N_c$ is the critical number of atoms for which the barrier
disappears and collapse is driven classically.  Macroscopic quantum
tunneling towards collapse can also occur through the barrier, as
sketched on the plot~\cite{ueda1998}.}
\label{fig:6}       
\end{figure}

Bright soliton solutions to the GPE with a constant external
potential $V(\vec{r})=V_0$ are unstable to collapse in three
dimensions.  In one dimension bright solitons are stable.  In two
dimensions they either collapse or expand indefinitely, depending on
the initial conditions and the strength of the effective nonlinear
coefficient. The precise balance between expansion and collapse is
known as the Townes soliton, or Townes
profile~\cite{chiao1964,moll2003}.  These now standard results for a
constant potential in one, two, and three dimensions are presented
rigorously in Sulem and Sulem~\cite{sulem1999}.  However, with the
addition of an external harmonic trapping potential, metastability
can be achieved in three dimensions.  This is easy to see by
considering the scaling of the three energy terms in the GPE. The
kinetic energy scales as $1/R^2$, where $R$ is the radius of a
single bright soliton.  The mean field energy scales as $-1/R^3$,
since the wavefunction is proportional to $R^{-3/2}$.  The potential
energy scales as $R^2$. Thus the additional contribution of the potential leads
to a metastable region.

In Fig.~\ref{fig:6} is shown a simple variational study for the
energy surface in the isotropic case, with $R$ taken as a
variational parameter for a Gaussian variational
ansatz~\cite{dodd1996b,dodd1996a,kagan1997,kagan1997a,perez1997,carr2002c,malomed2002}.
Since the state shown in Fig.~\ref{fig:6} is metastable, quantum
tunneling can cause the solution to tunnel through the barrier
towards $R=0$, i.e., collapse.  Ueda and Leggett~\cite{ueda1998}
derived an expression for the tunneling exponent based on a Gaussian
variational ansatz:
\begin{equation}
\frac{S^{\mathrm{B}}}{\hbar}\simeq 4.58 N
\left(1-\frac{N}{N_c}\right)^{5/4}
\end{equation}
where the tunneling rate is given by $\Gamma = A
\exp(-S^{\mathrm{B}}/\hbar)$ and $N_c$ is the critical number of
atoms past which the condensate loses metastability and becomes
unstable to collapse.  This is one of many instances in which
macroscopic quantum tunneling manifests in BECs, even within the
mean field
description~\cite{carr2002c,carr2004b,carr2005c,moiseyev2005}. One
must also make careful estimates of thermal fluctuations, which can
push the condensate up over the variational barrier.  A simple
estimate can be made by requiring that the thermal energy $k_B T$ be
much less than the difference between the energy of the metastable
state and that of the barrier peak, as sketched in Fig~\ref{fig:6}.

The tunneling barrier becomes large for condensates in prolate traps
with $\lambda\ll 1$.  The soliton then deforms from a spherical
shape to an elongated shape.  This is a simple experimental
signature of the effective dimensionality of the soliton. For oblate
traps, i.e., $\lambda\gg 1$, the soliton becomes two dimensional.
Then the condensate is either stable or unstable~\cite{perez1997}
within the radial degrees of freedom.  For sufficiently large
$|g_{\mathrm{eff}}|$ the condensate collapses; for smaller
$|g_{\mathrm{eff}}|$ its expansion is prevented by the external
harmonic potential.  The effect of the asymmetry $\lambda$ has been
studied both variationally and via numerical solution of the
GPE~\cite{perez1997,gammal2001,carr2002c}.

However, condensates in trapped BECs are always mathematically
metastable to three-dimensional collapse, whether the effective
dimensionality be one-, two-, or three-dimensional, due to quantum
tunneling.  It is simply that the tunneling time associated with
three-dimensional collapse becomes exponentially long; indeed, it is
so much longer than experimental lifetimes of 1 to 100 seconds that
it can be ignored.  BEC experiments are rife with such
metastabilities; for instance, the ground state of the kinds of
alkali metal gases used to make BECs is in fact a crystalline solid,
and the atomic gas is only in a metastable state, albeit long-lived.
In practice, we ignore all metastabilities not relevant to the time
scale of measurements, and assign an effective dimensionality to the
GPE to describe bright soliton properties.

Ignoring macroscopic quantum tunneling, the threshold for bright
soliton collapse can be determined by variational ansatz from the
mean field theory. This has an analytical expression in two special
cases, both of interest for BECs.  For an isotropic or nearly
isotropic
condensate~\cite{dalfovo1999,dodd1996b,dodd1996a,perez1997}, the
critical number of atoms is
\begin{equation}
N_c=0.6501\frac{\bar{\ell}}{|a|} \label{eqn:critical}
\end{equation}
where $\bar{\ell}\equiv (\ell_r^2\ell_z)^{1/3}$ is the geometric
mean of the harmonic oscillator lengths.  For a condensate which is
confined only in the radial direction, $N_c=0.7598
\,\ell_r/|a|$~\cite{carr2002c}. This case is especially interesting
as it corresponds to the propagation of a bright soliton in a
waveguide. Since bright solitons are themselves BECs which self-cool
to zero temperature~\cite{carr2002c}, they have been suggested as
carriers of information in atom circuits on a chip.  We note that
numerical studies of the GPE show that the actual critical number is
shifted by 10 to 20\% as compared to the variational result; this
can be incorporated by simply shifting the constant prefactor.

Lastly, although we have focused on the mean field and its linear
perturbations as described by the GPE and BDGE, this picture is
inadequate for describing the dynamics of attractive BECs past the
collapse threshold.  The essential reason is that the density
becomes so large that $\sqrt{n|a|^3}\sim 1$, where $n$ is the number
density and $a$ the scattering length.  The mean field theory, which
relies on a diluteness approximation~\cite{dalfovo1999,leggett2001},
necessarily breaks down at this point.  Nevertheless, a number of
attempts have been made to describe collapse dynamics with mean
field theories. For instance, some authors have modified the GPE by
adding an effective loss rate due to three-body
recombination~\cite{kagan1997,saito2002,gawryluk06:Talbot}.  Other
authors have considered a generalized time-dependent
Hartree-Fock-Boguliubov (HFB) theory which couples the mean atomic
field to a mean molecular field as well as normal and anomalous
atomic quantum fluctuations~\cite{milstein2003b,wuster2005}. A
recent extensive study of W{\"u}ster {\it et al.} \cite{wuester06ep}
has focused on reproducing the time of onset of collapse found in
the experiment of Ref.~\cite{donley2001} comparing simulations using
the GPE, HFB, and the stochastic truncated Wigner approximation
method for including the effects of quantum fluctuations. The
conclusion of this study was that the effect of quantum
fluctuations, as compared to GPE simulations, was small and could
not explain the discrepancies between the time scales found in the
simulations and the significantly faster collapse times seen in the
experiment.

\subsection{Bright Soliton Engineering: Pulsed Atom Lasers and Other Applications}
\label{ssec:atomlaser}

An area that has only begun to be explored experimentally is the
many regimes in which bright solitons are stable.  As pointed out in
Sec.~\ref{ssec:ims}, bright solitons are always metastable in
Bose-Einstein condensates, due to the geometries in which they are
made.  However, the instability times due to quantum tunneling can
be much longer than the lifetime of experiments.  All of the
``stable'' applications of BECs which are discussed in the
subsections~\ref{ssec:atomlaser}-\ref{ssec:brightring} are therefore
technically unstable; however, we use the term \emph{experimental
stability} to emphasize that from the point of view of measurement
they are stable.

\begin{figure}[t]
\center
\includegraphics[width=0.95\textwidth]{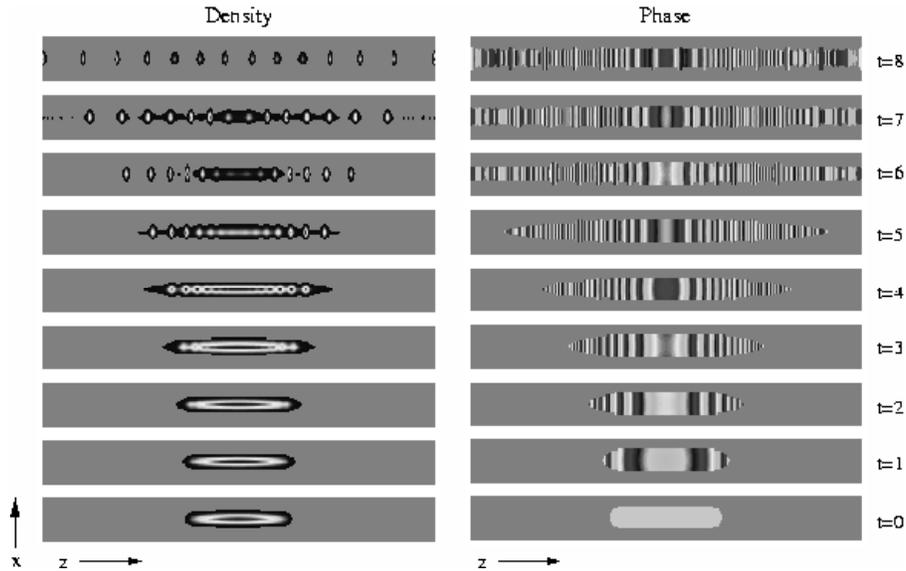}
\caption{{\it Pulsed atom soliton laser}: An initial state created
by changing the scattering length of a condensate from large and
positive to small and negative and then projecting the condensate on
to an expulsive harmonic potential results in spontaneous
modulational instability and a series of phase-coherent pulses, or
``mini-BECs.''  Shown are the evolution of the density and phase
along a two dimensional cut at $y=0$.  A set of well-defined
solitonic pulses is evident in the latest (top) panel. The strong
variations in the phase at late times is due to the high momentum of
the solitons caused by the expulsive harmonic potential. Note that
the phase is shown on the color circle, {\it i.e.}, modulo $2\pi$,
while the density is in arbitrary relative units rescaled for each
plot. For $N=10^4$ atoms, $a=-3 a_0$, and a trap geometry of
$\omega_{\rho}=2\pi\times 2.44$ kHz, $\omega_{z}=2\pi i\times 2.26$
Hz, the time units are scaled to 22 ms and the spatial units to 10
$\mu$m.  Note that the aspect ratio of the plots showing a region of
0.822 by 153 length units was changed for visualization.}
\label{fig:7}       
\end{figure}

In one of the first experimental demonstrations of a bright soliton,
a train of nearly 3D bright solitons was created from an elongated
BEC via modulational instability~\cite{strecker2002}, as described
in Chapter II.  On the other hand, another experiment published
simultaneously~\cite{carr2002c} produced a single bright soliton
with a weakly \emph{expulsive} harmonic potential in the $z$
direction, i.e., a harmonic trap turned upside down; the radial
harmonic trap was kept quite strong, so that the soliton propagated
down a waveguide.  The expulsive potential was then used to push the
soliton along and accelerate its dynamics.  A combination of these
two experimental techniques leads to a \emph{pulsed atom soliton
laser} as follows~\cite{carr2004h}. The large repulsive scattering
length of an initially highly elongated BEC is suddenly tuned small
and negative with a Feshbach resonance.  At the same time, the trap
is flipped over in the $z$ direction, i.e., $\omega_z\rightarrow
i\omega_z$.  The subsequent nonlinear evolution of the wavefunction
creates a series of pulses via modulational instability seeded by
linear interference fringes according to the Feynman
propagator~\cite{carr2004c}. These self-cooling ``mini-BECs'' each
contain on the order of $10^3$ to $10^4$ atoms. They are prevented
from overlapping, and thereby collapsing, by the expulsive harmonic
potential, and maintain their phase coherence over 500 ms.  This
sequence of events is illustrated in Fig.~\ref{fig:7}.

Improvements on this design have since been suggested in which many
more laser pulses can be produced from a better controlled
reservoir. In Chen and Malomed~\cite{chenPY2005}, a dual-core
approach is used to produce a matter-wave soliton laser from
attractive BECs. Two elongated quasi-one-dimensional condensates, or
``cores'' are laid side by side. The first condensate serves as a
reservoir for the second via macroscopic quantum tunneling.  The
scattering length is small and positive in the first condensate, and
small and negative in the second.  Bright solitons form in the
second condensate, and are emitted through a semi-transparent
barrier at one end.

Carpentier and Michinel~\cite{carpentier2006a} investigate this idea
in much greater detail by considering many possible spatial
variations of the scattering length to maximize output and control
over pulse size and velocity. In a second paper, they use a similar
idea to create a bright soliton accelerator in a ring-shaped trap,
similar to the ``nevatron'' already realized experimentally with
repulsive BECs~\cite{sauer2001}, but with many advantages over the
first demonstration~\cite{carpentier2006b}.  A \emph{temporal}
variation of the scattering length has also been used to engineer
bright solitons.  For instance, a bright soliton in free
three-dimensional space can be stabilized by rapidly oscillating the
scattering length from positive to
negative~\cite{saito2003,saito2004b,adhikari2004,abdullaev2003}.

We would like to point out that even though the ``engineering''
examples we have cited in this section do not take advantage of the
principles of quantum mechanical superposition or entanglement,
nevertheless the nonlinear effects which are key to their operation
result from averaging over a quantum many body wavefunction.
Moreover, quantum fluctuations must be considered in any serious
attempt at designing a pulsed atom laser and other such devices.
Therefore, they can be considered as examples of quantum
engineering.

\subsection{Solitons in a Thermal Bath}

The study of thermal effects on bright matter-wave solitons is
highly relevant in light of potential applications of solitons, and
has just begun to be explored.  The multidimensional aspects of
bright solitons in a BEC are very important in this context.  The
microscopic interactions between solitons and independent thermal
particles are described by the BDGE to lowest order in $1/N$, where
$N$ is the number of particles in the soliton.  In one dimension the
scattering problem of a single-particle with a soliton can be solved
exactly in the BDGE \cite{sinha2006} and the full quantum field
theory~\cite{mcguire1964}. It is found that the scattering of
thermal particles on the soliton is {\em reflectionless}, i.e., the
transmission coefficient is unity, which is a consequence of the
integrable nature of the nonlinear Schr{\"o}dinger (NLS) equation.
This is a very useful property for possible applications of bright
solitons in high-precision interferometry.

However, the extent into transverse dimensions that solitons have in
a wave-guide geometry breaks the integrability of the NLS equation
and allows for a finite reflection probability of scattering thermal
particles. During such reflection events, momentum is transferred
from the thermal particle to the soliton, which affects the
soliton's center-of-mass motion.  Therefore, a soliton immersed in a
thermal cloud can experience diffusive motion or be subject to a
frictional force when it is moving with a relative velocity to the
background.  In Ref.~\cite{sinha2006}, the friction and diffusion
coefficients were determined based on a calculation of the
reflection probability of thermal particles scattering off a
soliton.  Other approaches to describing the interaction based on
the Hartree-Fock-Bogoliubov formalism can potentially treat the
nonlinear coupled dynamics of the thermal cloud and the BEC. Studies
in this direction are reported in
Refs.~\cite{buljan2005,merhasin2006}.

\subsection{Soliton-Soliton Interactions}

Soliton-soliton interactions in one dimension have been described
completely and analytically by Gordon~\cite{gordon1983}.  They are
perfectly elastic.  However, new features arise in trapped BECs. The
imposition of a trapping potential can lead to chaotic dynamics for
three or more solitons, even in one dimension~\cite{martin2007}.
Higher-dimensional effects can lead to inelastic
collisions~\cite{carr2004h,parker2006,khaykovich2006}. The subject
of collisions of bright solitons in trapped BECs is only just
beginning to be studied.

The essential effects of higher dimensionality are as follows.  When
two identical bright solitons overlap they double their number of
atoms $N$.  Thus it is possible for $N$ to be temporarily greater
than $N_c$, as defined in Eq.~(\ref{eqn:critical}).  The time for
collapse to occur can be estimated from $g_{\mathrm{eff}}$.
Elasticity is then a question of whether or not the two solitons
spend enough time overlapping to undergo collapse or partial
collapse, at least within the mean field picture of the GPE.  This
is determined by their relative velocity~\cite{parker2006}.  An
additional factor is their relative phase and amplitude.  In one
dimension, a relative amplitude difference is equivalent to a phase
difference~\cite{gordon1983}. If the phase difference $\Delta\phi$
satisfies $\pi/2 \leq \Delta\phi \leq 3\pi/2$ then the solitons can
never overlap. On the other hand, if their phase difference
satisfies $-\pi/2 < \Delta\phi < \pi/2$, then partial overlap
occurs, with full overlap for $\Delta\phi=0$~\cite{carr2001a}.
Initial studies indicate that the situation is vastly more complex
in higher dimensions.  For example, bright solitons can collide so
inelastically that they ``stick,'' releasing excess energy and
relative momentum by emitting a few particles, similar to the way
that a soliton in one dimension adjusts to its preferred shape and
thereby self-cools~\cite{satsuma1974,carr2002c}.

It has been suggested, based on initial experiments on bright
soliton trains, that bright soliton collisions can in fact lead to
annihilation~\cite{strecker2002}. Beyond the mean field theory, it
is known experimentally that there is a bounce from collapse, as
described in Sec.~\ref{ssec:ims}. Consider two solitons in a soliton
train, each with a number of atoms near the critical number and
therefore nearly three dimensional.  The harmonic trap and/or
initial conditions can drive them to overlap.  If they do so for a
sufficient period of time partial collapse occurs, leaving one
soliton behind.  The loss of atoms is not properly described by the
mean field theory.  Even if the solitons are initially arranged with
nodes between them, drift of relative phase due to quantum
fluctuations can eventually lead to their being able to overlap.
This is one explanation of the occasional disappearance of a member
of the soliton trains of Ref.~\cite{strecker2002}.

Therefore, in addition to the exploration of bright soliton
collisions within the three-dimensional mean field theory of the
GPE, the effects of finite temperature~\cite{jackson2006} and higher
order quantum theories~\cite{milstein2003b} need to be considered as
well in order to model experimental dynamics.  The mean field theory
can only provide, at best, the threshold for collapse-related
effects.  This remains an important open problem for theorists to
address.

\subsection{Bright Ring Solitons and Quantum Vortices}
\label{ssec:brightring}

The attractive analog of a vortex in a repulsive BEC in free space
is in fact a bright ring soliton~\cite{carr2006c}.  This point is
illustrated in Fig.~\ref{fig:2}, where it can be seen that the
wavefunction approaches zero as $r\rightarrow\infty$, in
contradistinction to vortices in repulsive BECs where the
wavefunction approaches a non-zero constant.  The stability of
vortices in attractive BECs has been investigated
theoretically~\cite{ruprecht1995,dodd1996a,dodd1996b,pu1999} since
shortly after the experimental observation of a BEC.  However, no
experiment to date has tested theoretical predictions of bright ring
solitons.

Static studies, which consider stationary solutions of the GPE and
their linear perturbations as described by the BDGE, have found that
all bright ring solitons and their radial excitations are
unstable~\cite{pu1999,carr2006c}.  Initial studies predicted an
enhanced critical number over a bright soliton, and therefore
\emph{enhanced} stability. However, these studies considered only
radial collapse. In fact, bright ring solitons are azimuthally
unstable, as later analysis with the BDGE showed~\cite{pu1999}.
Although we have avoided significant analytical description thus far
in this review, it is useful to state the form of the BDGE for a
centrally located axisymmetric vortex~\cite{svidzinsky1998} in an
effectively two-dimensional condensate. We first transform the
Boguliubov amplitudes $u$ and $v$ according to
\begin{equation}
\pmatrix{u( \vec{r}) \cr v(\vec{r})} =
\frac{e^{im\phi}}{\ell_r}\,\pmatrix{e^{iq\phi}\,\tilde
u_m(\tilde{r})\cr e^{-iq\phi}\,\tilde
v_m(\tilde{r})}\,,\label{eqn:uvq}
\end{equation}
where $\tilde{r}=\sqrt{x^2+y^2}/\ell_r$ and we neglect perturbations
in the $z$ direction, considering only a strongly oblate trap for
simplicity.  Equation~(\ref{eqn:uvq}) represents a partial wave of
angular momentum $m$ relative to a condensate with a vortex of
winding number $q$. Then in harmonic oscillator units the BDGE
become
\begin{eqnarray}
\mathcal{L}_{+}\tilde{u}_m-\tilde{g}_{\mathrm{eff}}|\tilde{f}_q|^2\tilde{v}_m=\frac{\Omega_m}{\omega}\tilde{u}_m,
\label{eqn:bog3}\\
\mathcal{L}_{-}\tilde{v}_m-\tilde{g}_{\mathrm{eff}}|\tilde{f}_q|^2\tilde{u}_m=-\frac{\Omega_m}{\omega}\tilde{v}_m,
\label{eqn:bog4}
\end{eqnarray}
where
\begin{equation}
\mathcal{L}_{\pm} \equiv
-\frac{1}{2}\left(\frac{\partial^2}{\partial
\tilde{r}^2}+\frac{1}{\tilde{r}}\frac{\partial}{\partial \tilde{r}}
- \frac{(q\pm m)^2}{\tilde{r}^2}-\tilde{r}^2\right)
+2\tilde{g}_{\mathrm{eff}}|\tilde{f}_m|^2-\tilde{\mu}
\end{equation}
Here $\Omega_m$ are the eigenvalues for Boguliubov modes with
angular momentum $m$ and $\tilde{f}_q$ is the radial portion of
condensate wavefunction with winding number $q$; the tildes
throughout these rescaled BDGE indicate harmonic oscillator units.
The different centrifugal barriers inherent in $\mathcal{L}_{\pm}$
show that the two amplitudes behave differently near the origin,
with $\tilde u_m \propto \tilde{r}^{|m+q|} $ and $\tilde v_m\propto
\tilde{r}^{|m-q|}$ as $\tilde{r}\to 0$.  Note that $\tilde{f}_q$ is
normalized to unity.

We briefly highlight the results of studies of
Eqs.~(\ref{eqn:bog3})-(\ref{eqn:bog4}) for
$\tilde{g}_{\mathrm{eff}}<0$. The dominant instability mode is
quadrupolar for $m=1$, a single bright soliton, and small
$\tilde{g}_{\mathrm{eff}}$. The linear instability time is given by
$T_m=2\pi/\mathrm{Im}\Omega_m$. This time can be much longer than
experiments, so that bright ring solitons do indeed have the
possibility of being observed.

While multiple dark ring solitons consist of nested nodal rings,
separated by regions of non-zero density and constant phase,
multiple bright ring solitons consist of multiple rings of non-zero
density and constant phase separated by nodes.  These multiple
bright ring soliton solutions are in fact radial excitations of a
vortex in an attractive BEC.  In free space, there are a denumerably
infinite number of such states, which form an excited state spectrum
of bright ring solitons in two dimensions for fixed winding number
under the constraint that the wavefunction approach zero as the
radial coordinate approaches infinity.  This result has been
formally proved for winding number zero for the Townes
soliton~\cite{sulem1999} and numerically demonstrated for non-zero
winding number~\cite{carr2006c}.  The first excited state for
winding number $m=1$ is shown in Fig.~\ref{fig:2}(c).  This is in
contrast to dark ring solitons, which only appear in an infinite
number of concentric rings in free space, as shown in
Fig.~\ref{fig:2}(d).

\begin{figure}
\center
\includegraphics[width=0.95\textwidth]{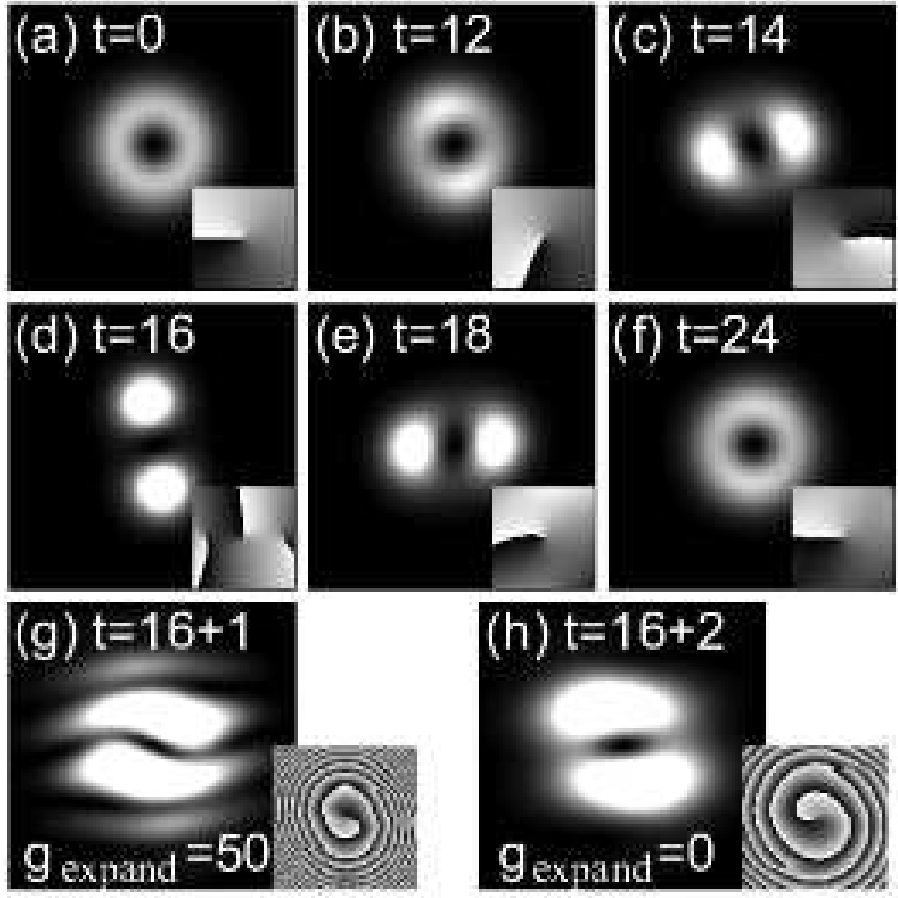}
\caption{{\it Nonlinear dynamical split-merge cycle of bright ring
soliton.} (a)-(f) Shown is the time evolution of the density profile
in two dimensions.  The insets present grey-scale plots of the phase
modulo $2\pi$. Panels (g) and (h) show what occurs when the trap is
switched off and the condensate is allowed to expand with (g) large
positive scattering length and (h) zero scattering length. This is a
common experimental technique to magnify condensate features too
small to otherwise resolve.}
\label{fig:8}       
\end{figure}

It is has been shown that a similar sequence of radially excited
states of attractive vortices occurs in a harmonic trap, and that,
for sufficiently small $\tilde{g}_{\mathrm{eff}}$, their instability
times can be long compared to experiments~\cite{carr2006c}.

Nonlinear dynamical studies of trapped bright ring solitons have
found cyclical behavior in their azimuthal break-up, among other
intriguing behaviors~\cite{saito2002b,saito2004}.  In
Fig.~\ref{fig:8} is shown one example (reproduced with permission of
the authors~\cite{saito2002b}). The initial state is a bright ring
soliton plus a small symmetry-breaking aziumthal perturbation. The
ring splits via the quadrupole instability into two density peaks,
i.e., two bright solitons.  The soliton pair rotates around the
origin, then recombines to reform the original ring.  Panels (g) and
(h) suggest the possibility of observation by sudden switching of
the scattering length and expansion of the condensate, a common
experimental technique. We note that parallels have been suggested
in multi-component BECs as well~\cite{garciaripoll2000}.

\section{Conclusion and Acknowledgements}

We have described a few of the many manifestations of soliton-like
phenomena in Bose-Einstein condensates in two and three dimensions.
A brief list for repulsive nonlinearity includes dark band solitons,
dark planar solitons, dark ring solitons, spherical shell solitons,
families of solitary waves, and skyrmions and vortex textures; while
for attractive nonlinearity one finds metastable bright solitons,
quantum tunneling and quantum evaporation of bright solitons, pulsed
atom soliton lasers, bright ring solitons, and the split-merge
cycle.  We think some of the most exciting outstanding problems in
this field are the higher order quantum and thermal effects on
solitonic phenomena, as we have indicated sporadically throughout
our discussion.

There is great deal both in and beyond our over one hundred
references that we have been unable to cover in the space allotted;
we sincerely hope we have not offended the many investigators whose
work we have not been able to include.

The authors would like to thank the editors of this book, who have
done a wonderful job facilitating a much-needed review of nonlinear
phenomena in BECs.  We thank our mentor William Reinhardt, who many
years ago led us both in the direction of solitons and BECs as a
graduate student and post-doctoral fellow, respectively.  LDC thanks
Charles Clark for many useful discussions and scientific
collaborations which led to writing this chapter. LDC gratefully
acknowledges the National Science Foundation for continuing support.


\end{document}